\documentclass[twocolumn,prx,floatfix,lengthcheck,citeautoscript,superscriptaddress,longbibliography]{revtex4-2}

\usepackage[normalem]{ulem}
\usepackage{amssymb,amsfonts,amsmath,amsthm}
\usepackage{graphicx}
\usepackage[english]{babel}
\usepackage[latin1]{inputenc}
\usepackage[pdftex,colorlinks=true,pdfstartview=FitH,linkcolor=blue,citecolor=blue,urlcolor=blue]{hyperref}

\usepackage{bm}
\usepackage{float}
\usepackage{mathtools}
\usepackage{color,soul}
\usepackage{times}
\usepackage{upgreek}
\usepackage{MnSymbol}
\usepackage{verbatim}
\usepackage[bottom]{footmisc}
\usepackage{bbold}

\usepackage[dvipsnames,svgnames,table]{xcolor}
\usepackage{hyperref}
\usepackage{fancyhdr}
\usepackage[english]{babel}
\usepackage[caption=false]{subfig}

\hypersetup{
pdfstartview={FitH},
colorlinks=true,    
linkcolor=NavyBlue, 
citecolor=Blue,   
filecolor=NavyBlue, 
urlcolor=NavyBlue   
}

\newcommand{\bea}{\begin{eqnarray}}
\newcommand{\eea}{\end{eqnarray}}
\newcommand{\be}{\begin{equation}}
\newcommand{\ee}{\end{equation}}

\newcommand{\sredout}[1]{%
  \ifmmode
\text{\color{red}\sout{\ensuremath{#1}}}
  \else {\color{red}\sout{#1}}%
\fi
}

\begin{document}
\title{
Synchronization in coherently and dissipatively coupled spinor polariton time crystals
}

\author{I. Carraro-Haddad}
\affiliation{Centro At{\'{o}}mico Bariloche and Instituto Balseiro,
Comisi\'on Nacional de Energ\'{\i}a At\'omica (CNEA)- Universidad Nacional de Cuyo (UNCUYO), 8400 Bariloche, Argentina.}
\affiliation{Instituto de Nanociencia y Nanotecnolog\'{i}a (INN-Bariloche), Consejo Nacional de Investigaciones Cient\'{\i}ficas y T\'ecnicas (CONICET), Argentina.}

\author{I. A. Ramos-P\'erez}
\affiliation{Centro At{\'{o}}mico Bariloche and Instituto Balseiro,
Comisi\'on Nacional de Energ\'{\i}a At\'omica (CNEA)- Universidad Nacional de Cuyo (UNCUYO), 8400 Bariloche, Argentina.}
\affiliation{Instituto de Nanociencia y Nanotecnolog\'{i}a (INN-Bariloche), Consejo Nacional de Investigaciones Cient\'{\i}ficas y T\'ecnicas (CONICET), Argentina.}

\author{G. Usaj}
\affiliation{Centro At{\'{o}}mico Bariloche and Instituto Balseiro,
Comisi\'on Nacional de Energ\'{\i}a At\'omica (CNEA)- Universidad Nacional de Cuyo (UNCUYO), 8400 Bariloche, Argentina.}
\affiliation{Instituto de Nanociencia y Nanotecnolog\'{i}a (INN-Bariloche), Consejo Nacional de Investigaciones Cient\'{\i}ficas y T\'ecnicas (CONICET), Argentina.}


\author{A. Bruchhausen}
\affiliation{Centro At{\'{o}}mico Bariloche and Instituto Balseiro,
Comisi\'on Nacional de Energ\'{\i}a At\'omica (CNEA)- Universidad Nacional de Cuyo (UNCUYO), 8400 Bariloche, Argentina.}
\affiliation{Instituto de Nanociencia y Nanotecnolog\'{i}a (INN-Bariloche), Consejo Nacional de Investigaciones Cient\'{\i}ficas y T\'ecnicas (CONICET), Argentina.}

\author{K. Biermann}
\affiliation{Paul-Drude-Institut f\"{u}r Festk\"{o}rperelektronik, Leibniz-Institut im Forschungsverbund Berlin e.V., Hausvogteiplatz 5-7,\\ 10117 Berlin, Germany.}

\author{P.~V. Santos}
\affiliation{Paul-Drude-Institut f\"{u}r Festk\"{o}rperelektronik, Leibniz-Institut im Forschungsverbund Berlin e.V., Hausvogteiplatz 5-7,\\ 10117 Berlin, Germany.}

\author{A.~S. Kuznetsov}
\email[Corresponding author, e-mail: ]{kuznetsov@pdi-berlin.de}
\affiliation{Paul-Drude-Institut f\"{u}r Festk\"{o}rperelektronik, Leibniz-Institut im Forschungsverbund Berlin e.V., Hausvogteiplatz 5-7,\\ 10117 Berlin, Germany.}

\author{A.~A. Reynoso}
\email[Corresponding author, e-mail: ]{a.a.reynoso@gmail.com}
\affiliation{Centro At{\'{o}}mico Bariloche and Instituto Balseiro,
Comisi\'on Nacional de Energ\'{\i}a At\'omica (CNEA)- Universidad Nacional de Cuyo (UNCUYO), 8400 Bariloche, Argentina.}
\affiliation{Instituto de Nanociencia y Nanotecnolog\'{i}a (INN-Bariloche), Consejo Nacional de Investigaciones Cient\'{\i}ficas y T\'ecnicas (CONICET), Argentina.}

\author{A. Fainstein}
\email[Corresponding author, e-mail: ]{alejandro.fainstein@ib.edu.ar}
\affiliation{Centro At{\'{o}}mico Bariloche and Instituto Balseiro,
Comisi\'on Nacional de Energ\'{\i}a At\'omica (CNEA)- Universidad Nacional de Cuyo (UNCUYO), 8400 Bariloche, Argentina.}
\affiliation{Instituto de Nanociencia y Nanotecnolog\'{i}a (INN-Bariloche), Consejo Nacional de Investigaciones Cient\'{\i}ficas y T\'ecnicas (CONICET), Argentina.}

\date{\today}

\begin{abstract}
{The spinor degree of freedom associated to exciton-polariton condensates can spontaneously self-oscillate breaking time translation symmetry, thus showing a continuous time-crystal (CTC) behavior. An open question in such driven-dissipative and non-linear quantum open systems is what happens when CTCs are brought together to interact. Here we experimentally study polariton condensates in coupled traps, evidencing mutual induction and synchronization of the pseudospin temporal GHz dynamics in the CTC phase. The individual and relative orientation of the (limit cycle) precessing pseudospins can be tuned by the optical excitation power, displaying both ferro and anti-ferro dynamical configurations. We theoretically show that the exciton reservoir, and both the coherent and long-range dissipative inter-trap coupling, play important roles in the CTC dynamics. The investigation of time-broken symmetry is thus extended here to more complex non-hermitian systems opening the path to study self-sustained collective dynamics in lattices of non-linear quantum condensates.
}
\end{abstract}
\maketitle

Time crystals identify quantum systems exhibiting spontaneous breaking of the time translation symmetry (TTS)~\cite{Wilczek2012,Bruno2013,Sacha2018}. It has been shown that TTS can naturally break in an open quantum system where losses are compensated through continuous pumping~ \cite{Kessler2021,Autti2018,Gong2018,Tucker2018,Zhu2019,Buca2019,Lazarides2020}. These systems are termed ``continuous time crystals'' (CTCs), as their Hamiltonians are time-independent, in contrast with  TTS broken in periodically-driven many-body Hamiltonians~\cite{Sacha2018,Else2020}. The latter are characterized by stable states with periods distinct from the drive and called Floquet or ``discrete" time crystals (DTCs)~\cite{Kessler2021,Autti2018,Zhang2017,Taheri2022,Frey2022}. Experimental realizations of CTCs have been recently reported based on atoms confined in laser traps ~\cite{Kessler2019,Kongkhambut2022}, in magnon condensates in superfluid $^3$He~\cite{Autti2018}, for  optically excited nuclear spins in semiconductors~\cite{Greilich2024}, and also involving the limit-cycle Larmor-like precession of the pseudospin degree of freedom of exciton-polariton condensates~\cite{Carraro2024}.

As the identification and experimental study of such stable CTC dynamics in non-Hermitian and non-linear open quantum systems develops, fundamental questions related to the coupling of time crystals gain relevance. Theoretically, \textit{boundary} time crystals~\cite{Iemini2018} have been addressed, where continuous TTS breaking occurs in a macroscopic fraction of a many-body quantum system. In addition, the idea of \textit{seeding} of crystallization in time was recently introduced~\cite{Hajdusek2022}, where  a subsystem acts as a nucleation center to induce TTS breaking across an entire ensemble. 
When a larger system induces the emergence of a time-periodic steady state in another, what are the possible frequencies and phase patterns? Do such CTCs subsystems synchronize? Dissipative spin systems as described by the Rabi-Hubbard model have been used to address these questions~\cite{Schiro2016}. Experimentally, the coupling of TCs occurring at the boundary and interior of a $^3$He superfluid has been reported in \cite{Autti2021}, with Josephson-like oscillations developing between the two coupled magnon condensates.

Here we show that the spinor degree of freedom in driven-dissipative exciton-polariton condensates of coupled engineered trap structures represent an ideal system to address these intriguing questions. We experimentally study the spectra and dynamics associated to the Larmor-like CTC precession of the pseudospin in condensates confined in close (separation $d \sim 1 \mu$m) and distant ($d \sim 15 \mu$m) micro-traps. It is shown that synchronization of both the condensate energies and of their pseudospin precession occur, with average orientation resembling both ferro and antiferro-like structures depending on the optical excitation power. Theoretical modeling allows to identify the role of the exciton reservoir, and of Josephson-like coherent and long-range dissipative couplings, on the observed CTC dynamics.

\begin{figure*}[t]
 \begin{center}
    \includegraphics[trim = 0mm 0mm 0mm 0mm,clip=true, keepaspectratio=true, width=2.0 \columnwidth, angle=0]{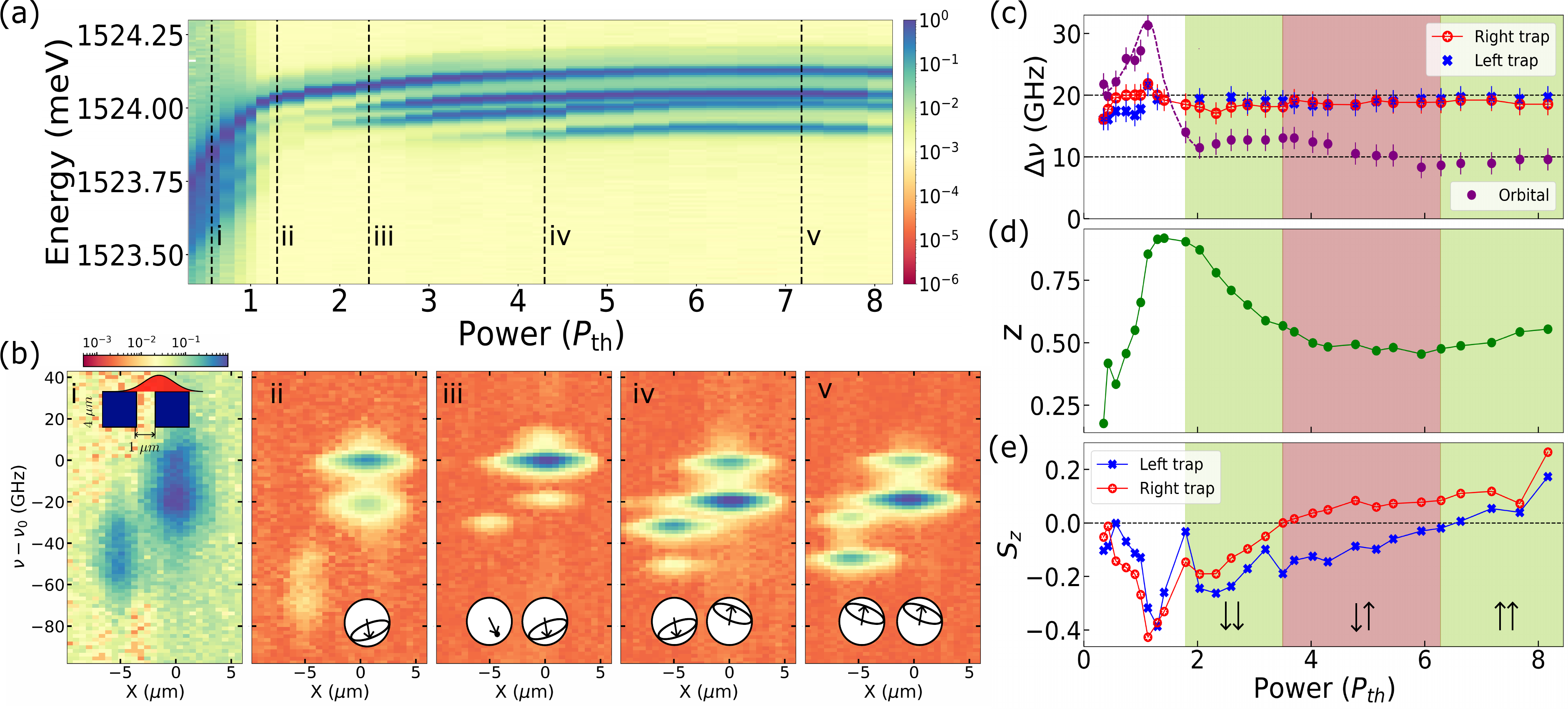}
\end{center}
\vspace{-0.8 cm}
\caption{\textbf{Spinor coupled time crystals with an internal clock.} 
Experiments on two $4 \times 4~\mu$m$^2$ square polariton traps, separated by a small $1~\mu$m barriers. A continuous wave linearly polarized pump laser non-resonantly injects free carriers with $P\mathrm{cw}$. These relax, forming an exciton reservoir mostly feeding one of the polariton traps. \textbf{(a)} Photoluminescence spectra integrated from the two traps for increasing non-resonant cw laser power (magnitude given in terms of the condensation threshold power  $P_\mathrm{th}$. \textbf{(b)} Spectrally-resolved photoluminescence images for the five powers indicated with vertical dashed lines in in \textbf{(a)}. A scheme of the two traps and and of the optical excitation configuration is shown in the left-most panel. At the bottom of panels ii-v a Bloch-sphere representation of the associated spinor solutions is shown (see text for details).
\textbf{(c)} Frequency splitting as a function of excitation power, between the spinor components in each trap (red and blue symbols) and between modes corresponding to the two different traps (black symbols). \textbf{(d)} Relative integrated intensity between modes of the right and left traps, $(I_\textrm{R}-I_\textrm{L})/(I_\textrm{R}+I_\textrm{L})$, as a function of excitation power. \textbf{(e)} Pseudospin $z$ component for the left and right traps, as a function of excitation power. Shaded colors represent alternating regions corresponding to parallel and anti-parallel oriention of the spinor orbits (identified with arrows in each region).
In panels \textbf{(c-e)} the excitation power is given in terms of the threshold power $P_\textrm{th}$. The vertical dashed lines indicate the powers at which pseudospin orientation reversal occurs. 
}
\label{Fig1}
\end{figure*}

\textit{Coupled polaromechanical continuous time crystals}.--- 
Our platform is based on exciton-polaritons (shortly, polaritons), quasiparticles resulting from the strong coupling between excitons and photons in GaAs/AlAs based semiconductor three dimensional traps. Polaritons are bosonic excitations with properties that follow from their two constituents, namely a small mass and a dissipative nature deriving from the photonic component, and mutual Coulomb interactions and efficient coupling to phonons due to the excitonic component ~\cite{Hartmann2006,Amo2009,CarusottoRMP2013,Rozas2014}.  
Under appropriate conditions (temperature and optical excitation power), polaritons transition to a non-equilibrium Bose-Einstein condensate (BEC) ~\cite{Kasprzak2006,Balili2007}. The energy levels of exciton-polaritons based on semiconductor materials can be engineered by making trapping potentials using different technologies, including deep etching, spatially modulated illumination, or cavity-spacer microstructuring~\cite{Schneider2016,Bajoni2008,Galbiati2012,Askitopoulos2013,Cristofolini2013,Alyatkin2021,Winkler2015,Kuznetsov2018}. 
In this way, individual traps, coupled Josephson-like junctions, or two-dimensional arrays, can be designed with tailored energy levels and couplings, and long spatial (tens of microns) and time (some ns) coherence.

The traps also confine $\sim 20$~GHz mechanical vibrations that very efficiently couple to polaritons through deformation potential interaction~\cite{Fainstein2013,Santos2023}. Due to this interaction, polariton condensates in two coupled traps with energy levels detuned at the phonon energy, drive mechanical self-oscillation (phonon lasing)~\cite{RMP,Chafatinos2020,Reynoso2022}. Conversely, this mechanical modulation of the polariton levels induces their mutual locking at energies that match with integer numbers of the phonon frequency~\cite{Chafatinos2023,Ramos2024}.
Recently, we showed that the pseudospin degree of freedom of the polariton ground state in an isolated square trap with a small rectangular distortion, can develop a non-linear self-sustained dynamics (a CTC) when subject to non-resonant continuous wave (cw) excitation~\cite{Nalitov2019,Carraro2024}. When the frequency of this  Larmor-like limit-cycle gets close to that of confined vibrations in the structure, it resonantly drives phonons and leads to coherent mechanical oscillation. As in the case of coupled traps mentioned above, these coherent mechanical oscillations in turn back-act locking the CTC frequency (a clock)~\cite{Carraro2024}. Here we use this platform to describe the resulting dynamics when spinor TCs in dimers and triads of traps become coupled. 

Figure~\ref{Fig1}(a) presents a color map of the excitation power dependence of the photoluminescence (PL) spectrum of a polariton condensate junction made of two square $4 \times 4\, \mu$m$^2$ traps separated by a $1 \mu$m thick barrier (powers are given in terms of the BEC threshold power $P_\mathrm{th}$). The structure has been fabricated by laterally micro-patterning the spacer region of an otherwise planar microcavity~\cite{Winkler2015,Kuznetsov2018} (see Methods). For these experiments the lower polariton band has a negative detuning of around $9$~meV (corresponding to $\sim 90\%$ photonic strength), with the polariton Rabi gap being $\Omega_\textrm{R} \sim 6$~meV and the junction barrier $\sim 7$~meV high. The Josephson coupling for this polariton junction has been calculated using a generalized Gross-Pitaevskii (gGP) modeling~\cite{Wouters2007} to be around $J ~ \sim 10$~GHz ($\sim 40 \mu$eV). Non-resonant cw optical excitation is used, with a slight elliptical polarization and focused to a $\sim 5\, \mu$m diameter Gaussian spot. 
The laser spot is positioned closer to the right trap, but with its Gaussian tale overlapping also with the one on the left (see the scheme in Figure~\ref{Fig1}(b)). Electron-hole pairs are thus optically injected at high energies, they subsequently loose energy (erasing any phase information set by the exciting laser), until a relaxation bottleneck leads to the build-up of an exciton reservoir with energy close to the polariton junction barrier. This reservoir is slightly polarized due to the excitation laser polarization, with unequal population of spin-up $n_{+}$ and spin-down $n_{-}$ excitons.  The reservoir acts as a bath, feeding the polariton ground state (GS). Quite notably, the spectra in Fig.~\ref{Fig1}(a) display four lines, evidencing that the two GS of the traps are not only coupled, but also split in their two pseudospin components with increasing power. We now discuss in detail the implications of this peculiar phenomenon.


Figures~\ref{Fig1}(b) shows spectrally resolved spatial images for some selected laser excitation powers indicated with vertical dashed lines in Fig.~\ref{Fig1}(a). Below threshold (i) a broad emission is detected from both traps, the one more strongly pumped is slightly blue-shifted due to interactions. When the excitation power increases, condensation is first observed in the pumped trap, evidenced by the line-narrowing and the increase in intensity (ii). 
The strong energy blue-shift of the traps is now evident, determined by polariton-polariton and polariton-reservoir exciton-mediated interactions. Notably, already at this pump power the ground state emission is split. These lines are due to the  $\sigma=\pm$ polarization degree of freedom of the condensate, directly linked to the spin of the excitons and to the polarization of the cavity photons through angular momentum selection rules~\cite{Shelykh2005,Ohadi2015,Gnusov2020,Siggurdsson2020}. As previously reported, the two observed split lines are mutually coherent, and result from a stable Larmor-like precession of the pseudospin~\cite{Nalitov2019,Carraro2024}. The splitting corresponds to the pseudospin precession frequency, and notably becomes locked to the cavity confined phonon frequency $\nu_\textrm{M} \sim 20$~GHz. On further increasing the excitation power (iii) polaritons also condense on the neighbor (left) trap, which however is mostly monochromatic (no pseudospin Larmor-like precession is evident). CTC behavior at  $\nu_\textrm{M} \sim 20$~GHz is also induced on the neighbor trap at higher powers (iv), situation at which both traps attain an energy separation that closely matches half the phonon frequency ($\nu_\textrm{M}/2 \sim 10$~GHz). At the highest illustrated power (v) both pseudospins are precessing in a stable CTC state. The relative intensity of the higher and lower energy components of each pseudospin emission is however reversed respect to the situation at lower powers. As we will further describe below, this reversed intensity distribution in the pseudospin doublet reflects a power induced flipping of the orientation of the pseudospin Larmor-like precession.
From these spectra additional information can be obtained, as presented in Figs.~\ref{Fig1}(c,d). Namely, 
the dependence with incident laser power of the frequency separation between the two components of each spinor, and between the states of both traps (c), and the relative integrated intensity of the right and left traps (d), $\Delta I/I_\mathrm{tot}=(I_\textrm{R}-I_\textrm{L})/(I_\textrm{R}+I_\textrm{L})$.

Several important results can be extracted from these data, namely: $(i)$ the splitting of the pseudospin components belonging to both traps starts somewhere around $\Delta_\mathrm{s} \sim 15$~GHz at $P_\mathrm{th}$, and rapidly stabilizes and locks at $\Delta_\mathrm{s} \sim \nu_\textrm{M} = 20$~GHz; $(ii)$ The detuning between traps, on the other hand, rapidly grows up to about $\Delta_\mathrm{J} \sim 30$~GHz above threshold, but around $P \sim 2P_\mathrm{th}$ the pumped trap saturates its blueshift and the left one catches up stabilizing around $\Delta_\mathrm{J} \sim \nu_\textrm{M}/2 = 10$~GHz (see Fig.\ref{Fig1}(c)); and $(iii)$, the intensity of the pumped trap rapidly increases above threshold, while the left trap shows an abrupt and strong increase of its population coincident with $P \sim 2P_\mathrm{th}$ (Fig.\ref{Fig1}(d)). The onset of this transfer of relative population between traps also coincides with the stabilization of their detuning around $\sim 10$~GHz. The spectra shows that the pseudospin splitting of the left trap becomes apparent when the two traps asynchronously lock at $\nu_\textrm{M}/2 \sim 10$~GHz. 
Moreover, the spectra display, in addition to the main lines, weak sidebands visible at higher frequencies and separated by $\nu_\mathrm{m}/2$.  These sidebands confirm the presence of a dynamical state, and are indicative of period doubling referred to the phonon modulation~\cite{Carraro2024}.


\textit{Experimental evidence of magnetically ordered rotating spinor phases}.---It has been previously shown that above threshold a spatially trapped exciton-polariton condensate undergoes a spontaneous parity breaking bifurcation to a ferromagnetic state~\cite{Ohadi2015}. At this transition, the whole condensate spontaneously magnetizes and randomly adopts one of two elliptically polarized states with opposite handedness of polarization. This study was later extended to multiple coupled spinor polariton condensates, which were shown to span different (ferromagnetic (FM) or antiferromagnetic (AFM)) phases depending on the incident power and on the physical parameters when arranged in a square geometry~\cite{Ohadi2017,Sigurdsson2017}. In both studies, the observed polariton states correspond to stable fixed points, that is, time independent solutions characterized by a single spectral line~\cite{Carraro2024}. Here we study the case of coupled polariton pseudospins when TTS is broken and these have transitioned to oscillating limit-cycle solutions. As we have shown, these solutions correspond to superposition of states that are phase locked and that are typically accompanied by a frequency comb emission~\cite{Rayanov2015,Carraro2024}. 

Figure~\ref{Fig1}(e) shows the $z$-component of the spinor in each of the two traps, obtained by spatial filtering and collecting the two circular polarization components $I^{+,-}_\mathrm{L,R}$, namely,  $S^\mathrm{z}_\mathrm{L,R} = (I^{+}_\mathrm{L,R} - I^{-}_\mathrm{L,R})/(I^{+}_\mathrm{L,R} + I^{-}_\mathrm{L,R})$, where $L(R)$ stands for light collected from the left(right) trap. 
\begin{figure}[t]
    \centering 
    \includegraphics[trim = 0mm 0mm 0mm 0mm,clip=true, keepaspectratio=true, width=1\columnwidth,angle=0]{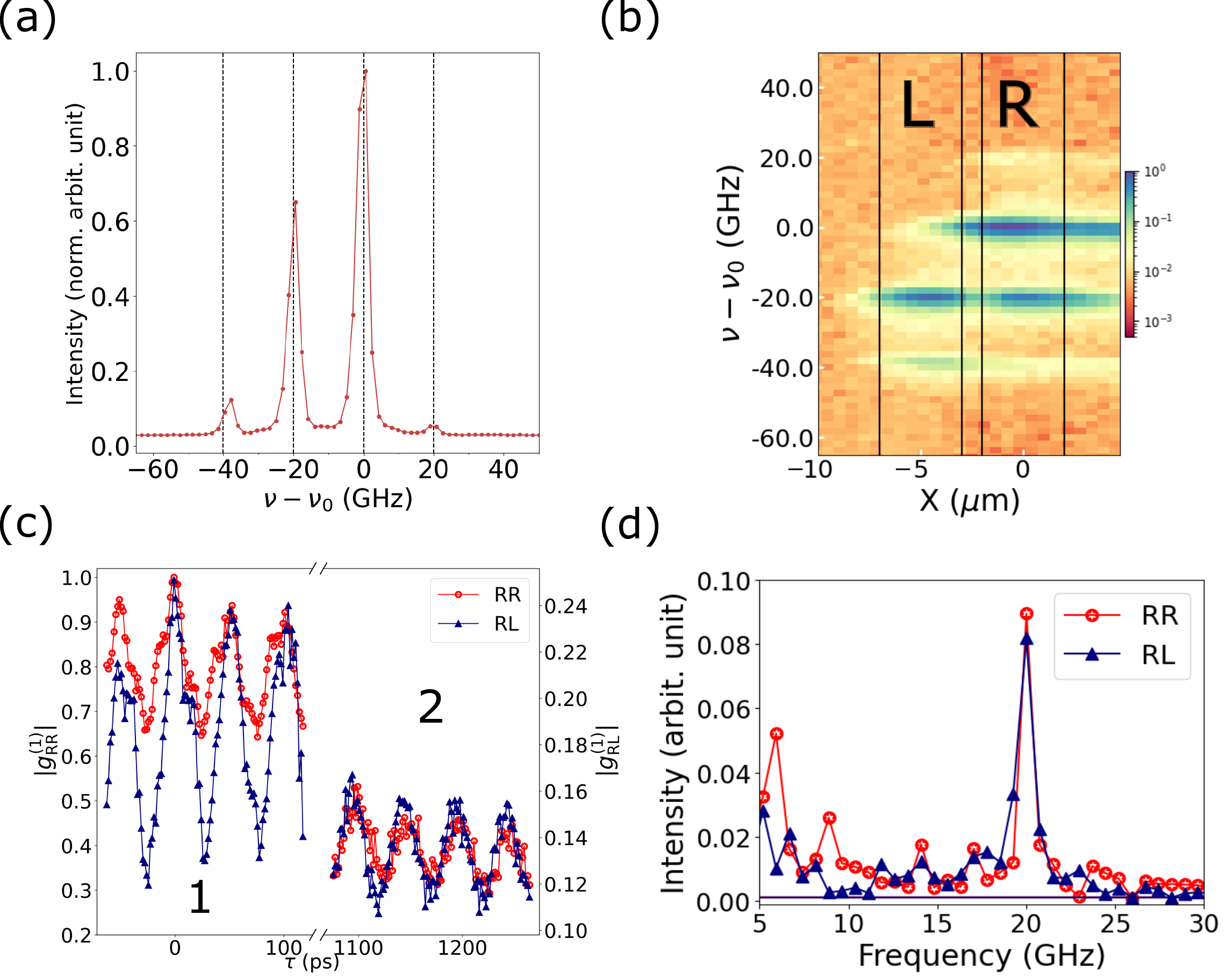}
    \caption{
    \textbf{Time dependence of the polaromechanical CTC molecule determined through $g^{(1)}(r,\tau)$.}     
     \textbf{(a)} Spectrum integrated from the whole structure, corresponding to two coupled $2 \times 2\mu$m$^2$ square traps separated by $2 \mu$m.  Frequencies are given respect to that of the most intense peak. Vertical lines are separated $\nu_\mathrm{M}=20$~GHz. \textbf{(b)}  Spectrally resolved spatial image corresponding to the spectrum in \textbf{(a)}.  \textbf{(c)} Time-dependent $g^{(1)}(r,\tau)$  shown for two situations, either ``direct'' with the light from the traps interfering with themselves (red dots), or ``crossed'' with the light from the right trap interfering with that of the left (blue triangles). Two time windows are shown, for delays $\tau$ around 0 and 1100~ps respectively. \textbf{(d)} Fourier transforms of the data in  \textbf{(c)}.
}   
    \label{Fig2}
\end{figure} 
At threshold the spinor of the two traps is tilted towards the `South' of the Bloch sphere  $S^\mathrm{z}_\mathrm{L,R}<0$, induced by the slight elliptical polarization of the pump. As power is increased, a successive flip of the spinor precession orbits occurs:  at $P \sim 4P_\mathrm{th}$ first the right trap spinor changes to $S^\mathrm{z}_\mathrm{R} > 0$ and then,  at $P \sim 6P_\mathrm{th}$, also $S^\mathrm{z}_\mathrm{L}> 0$. This shows that the orientation of the Larmor-like precession of the TCs in both traps can be controlled by the excitation power, so that first they are aligned ferromagnetically-like (pointing down), then antiferro-like, and finally ferro-like again (pointing up). 


\textit{Synchronization of TCs evidenced by the first order autocorrelation $g^\mathrm{(1)}(\tau)$}.---The observed locking of the pseudospin frequency detuning at   $\Delta_\mathrm{s} \sim \nu_\textrm{M} \sim 20$~GHz, and the emergence of sidebands, are evidence for the existence of a coherent time dynamics involving the polaritons~\cite{Carraro2024,Rayanov2015}. In addition, direct information of the time-dependence can be obtained through the time-resolved spatial first-order coherence function $g^{(1)}(r,\tau)$ (see Methods). This study, performed for two $2 \times 2\,\mu$m$^2$ square traps separated by $2\, \mu$m, is presented in Fig.~\ref{Fig2}. The corresponding spectrum and spectrally resolved spatial image are shown in Figs.~\ref{Fig2}(a) and (b), respectively (frequencies are given relative to the most intense line). The spatial image shows that two of the lines belonging to the individual traps synchronize [observed at $-20$~GHz in (a-b)]. The accompanying pseudospin doublet on each trap is shifted precisely $\Delta_\mathrm{s} = \nu_\textrm{M} \sim 20$~GHz, to lower (higher) energies in the left (right) trap. In addition, a weak sideband is observed at $\sim 20$~GHz. Similar to the case presented in Fig.~\ref{Fig1}, this spectrum evidences: (i) a Larmor-like precession of the pseudospins in the two coupled traps; (ii) that the precession frequency is locked at the phonon frequency; and (iii) that the detuning between traps becomes locked. 

Two situations for $g^{(1)}(r,\tau)$ are illustrated in Fig.~\ref{Fig2}(c), with the pattern obtained either with the light from the two traps interfering with themselves ($g^{(1)}_\mathrm{RR}$, red dots, left scale), or shifting the images on the CCD so that the light from the right trap interferes from that of the left one ($g^{(1)}_\mathrm{RL}$, blue triangles, right scale). Two time windows are shown, for time delays $\tau$ around $0$ [region (1)] and $1100$~ps [region (2)], respectively. The Fourier transforms of the data in (c) are presented in Fig.~\ref{Fig2}(d).   Temporal oscillations, that are persistent up to the longest measured delays ($\sim 1.3$~ns) and with a period that exactly matches the inverse of the phonon frequency $1/\nu_\mathrm{m} \sim 50$~ps, are clearly observed. They are on the other hand consistent with the spectral line-splitting as presented in Fig.~\ref{Fig2}(a). These results clearly establish that the spinors of the two condensates are oscillating in a stable Larmor-like precession, which is locked to the confined phonon frequency (a CTC with a mechanical clock, as described in Ref.~\onlinecite{Carraro2024}). Most notable, in addition, the fact that $g^{(1)}_\mathrm{RR}$ and $g^{(1)}_\mathrm{RL}$ coincide implies that the orbits are not only perfectly locked in frequency but also share the same phase. 

\begin{figure}[t]
    \centering 
    \includegraphics[trim = 0mm 0mm 0mm 0mm,clip=true, keepaspectratio=true, width=1\columnwidth,angle=0]{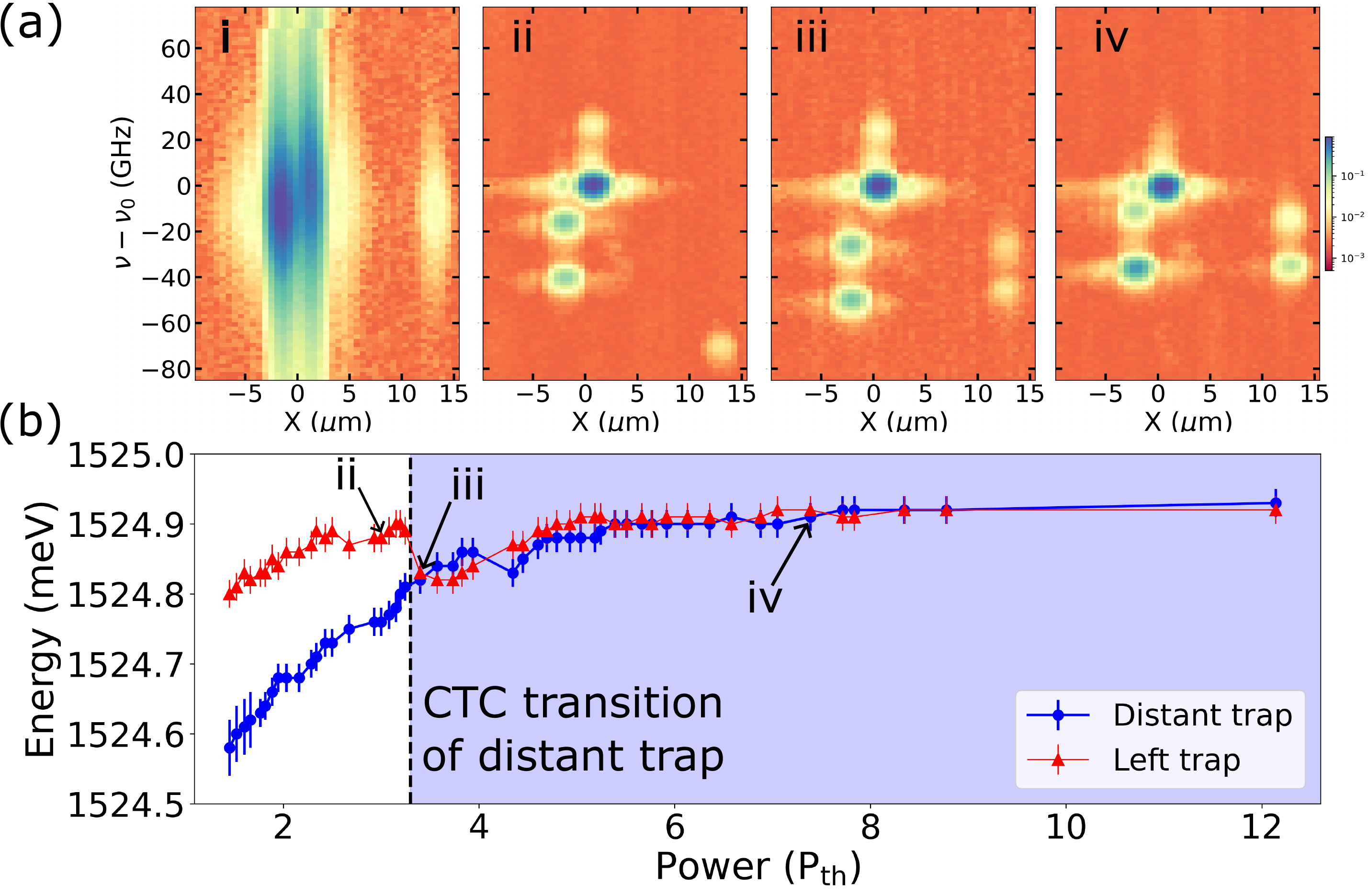}
    \caption{
        \textbf{Dissipative coupling of a TC to a distant condensate.}  
        \textbf{(a)} Spectrally resolved spatial images for varying incidence power for a structure consisting of two $2 \times 2 \mu$m$^2$ square traps separated by $2 \mu$m barriers (as the example in Fig.~\ref{Fig3}), but with in addition a distant identical trap separated $13 \mu$m from the center of the right (R) trap.  \textbf{(b)}Energy position as a function of incidence power (referred to $P_\mathrm{RM}$) for the high energy components of the left (L) and the distant (D) trap.  The blue shade indicates the region in which the L and D traps fully synchronize, and coincidentally the TC behavior is induced in D.
   }
     \label{Fig3}
\end{figure} 

\textit{Seeding of a distant time crystal: dissipative coupling}.---The experiments described so far demonstrate that the TC dynamics in the two traps becomes synchronized, that the energies of the polariton BECs lock at specific values related to the confined phonon frequency, and that the two phenomena are connected.  That is, not only the spinor modes split in the same amount ($\nu_\textrm{M} \sim 20$~GHz) and lock their phase, but also the condensate energies attain differences that match the phonon frequency or a fraction of it.  
To gain further clarity on the nature of the inter-trap coupling, Figure~\ref{Fig3} illustrates an experiment performed on a polariton `molecule', as described in the previous sections, consisting of two $2 \times 2\, \mu$m$^2$ square traps separated by $2 \,\mu$m barriers (as the example in Fig.~\ref{Fig2}), but with in addition a distant identical trap separated $13\, \mu$m from the center of the right (R) trap. The direct overlap between the ground states of the `molecule' traps and the distant traps is, in this case, negligible. As in the previous experiments, the excitation is non-resonant, with polarization slightly elliptical, and with the Gaussian spot $\sim 5\, \mu$m in diameter placed on top of the  ``molecule'' pair and shifted towards the right (R) trap. Figure~\ref{Fig3}(a) presents some selected spectrally resolved spatial images (power increasing from $i$ to $iv$), while Fig.~\ref{Fig3}(b) displays the energies as a function of incidence power (referred to $P_\mathrm{th}$) for the high energy components of the left (L) and the distant (D) trap.

The two closely coupled traps follow, as a function of increasing power, essentially the same pattern  as the case presented in Fig.~\ref{Fig1}. The distant trap, on the other hand: (i) is weakly observed below threshold; (ii) blue-shifts and it is seen as a single line at higher powers when the two pumped traps are clearly already in a coupled dynamical TC state; (iii) it suddenly develops a TC behavior when approaching the energy of the L trap around $P \sim 3.2 P_\mathrm{th}$, and locks both its splitting and its absolute energy with that of the L trap; and (iv) it continues fully synchronized to the L trap up to the highest pump powers ($P \sim 12 P_\mathrm{th}$). Notably, when this latter full TC and polariton energy synchronization is produced, the L trap reverses its orientation to $S^\mathrm{z}_\mathrm{R} > 0$ and the D trap is seeded with the same orientation (shaded region in Fig.~\ref{Fig3}(b)). Because of exciton-exciton repulsion, the exciton reservoir can attain at such high powers dimensions well larger than $10\, \mu$m, even if the spot is of only $\sim 5\, \mu$m. Thus, pumping of the D trap is no surprise. In this sense, the seeding of the TC behavior could occur for the D trap mediated by the reservoir dynamics~\cite{Carraro2024}. But its energy synchronization is, in view of the almost null trap wavefunction overlap, only possible if one allows for dissipative (radiative) coupling between the distant traps~\cite{Aleiner2012}. In what follows we will present a model to comprehensively describe the experimental observations. 

\textit{Theoretical model of coupled time crystals}.---The coupled TC dynamics investigated here arises from the interplay between driven$-$dissipative spinor polariton condensates, a structured exciton reservoir network, and optomechanical back$-$action mediated by cavity$-$phonon modes. Each condensate, located at trap $j=1,2$, supports two spin projections, $\sigma=\pm$, described by complex fields $\tilde{\Psi}_{j\sigma}(t)$. Their dynamics are governed by generalized spinor Gross$-$Pitaevskii equations (GPEs) that incorporate nonlinear interactions, coherent mixing between the two spin components, and either coherent tunneling or radiative coupling between the spatial sites. This phenomenological model captures the essential ingredients leading to the formation, frequency synchronization, mutual energy locking, and mechanical back$-$action of polariton TCs. The GPE reads
\begin{eqnarray}
\nonumber
i\hbar \frac{d\tilde{\Psi}_{j\sigma}}{dt} &=& (\hbar\omega_{j} + U_0 |\tilde{\Psi}_{j\sigma}|^2
+ U_0^\mathrm{R} \tilde{n}_{j\sigma}) \tilde{\Psi}_{j\sigma} \\
\nonumber
&& - J_{j}  \tilde{\Psi}_{j\bar{\sigma}}
+ \frac{i\hbar\gamma}{2}\left(\sum_i r_{ji}\tilde{n}_{i\sigma} - 1\right)\tilde{\Psi}_{j\sigma} \\
&& - (J_{\mathrm{12}}+ i J_{\mathrm{12},d}) \tilde{\Psi}_{3-j,\sigma}\,.
\label{eq:psij}
\end{eqnarray}

The terms in Eq.~\eqref{eq:psij} describe distinct physical processes. The first group sets the on$-$site energy of each trapped condensate, $\hbar\omega_j$, and includes its nonlinear renormalization through same$-$spin interactions, $U_0|\tilde{\Psi}_{j\sigma}|^2$, 
and the reservoir$-$induced blueshift $U_0^\mathrm{R} \tilde{n}_{j\sigma}$. The $J_j$ in each trap  accounts for intra-trap coherent mixing between the two spin components, which arises from a strain$-$induced splitting of the linearly polarized modes. The gain provided by the exciton reservoirs enters through the term proportional to $\sum_i r_{ji}\tilde{n}_{i\sigma}$. The gain matrix $r_{ji}$ allows reservoir excitations at site $i$ to feed the condensate at trap $j$. We choose $r_{11}=r_{22}=1$ for the feeding to be dominated by the local reservoir, while the nondiagonal contribution $r_{12}=r_{21} \ll r_{11}$  generates a dissipative coupling channel between the two condensates even in the absence of coherent tunneling. Finally, the term $(J_{12} + iJ_{12,d})\tilde{\Psi}_{3-j,\sigma}$ introduces an inter-trap coupling, which plays a central role in establishing mutual energy synchronization between the two trapped condensates. This coupling can have a coherent, Josephson-like, component, $J_{12}$, and/or a dissipative one, $J_{12,d}$.

The reservoirs evolve according to
\begin{equation}
\label{eq:ni}
\frac{d\tilde{n}_{i\sigma}}{dt}
= \gamma_{\mathrm{R}}\!\left[p_{i\sigma}
- \left(1 + \sum_j r_{ij}|\tilde{\Psi}_{j\sigma}|^2\right)\tilde{n}_{i\sigma}\right]\,,
\end{equation}
reflecting the competition between incoherent pumping and depletion via stimulated scattering into the condensates. Because the depletion rates depend on the occupations of both condensates through the matrix $r_{ij}$, the reservoirs mediate an effective non-local dissipative interaction. This coupling plays an important role in stabilizing both the amplitude and frequency of the two TC limit cycles and enables synchronization even when coherent tunneling is weak.

Here, to facilitate the comparison with the condensation threshold, the dynamical variables have been rescaled from their original versions ($\Psi_{j\sigma}$ and $n_{j\sigma}$) using the characteristic densities $n_0 = \gamma/R$ and $\rho_0 = \gamma_\mathrm{R}/R$, namely, $\tilde{\Psi}_{j\sigma} = \Psi_{j\sigma}/\sqrt{\rho_0}$ and $\tilde{n}_{j\sigma} = n_{j\sigma}/n_0$. In this convention, $R n_{j\sigma}$ represents the filling rate of the condensates from the reservoir and $\gamma_\mathrm{R}$ the decay rate of excitons. Likewise, the reservoir pump power has been normalized as $p_{j\sigma} = P_{j\sigma}/P_\mathrm{th}$, where $P_\mathrm{th} = \gamma\gamma_\mathrm{R}/R$ is the condensation threshold.

Beyond the optical and reservoir subsystems, the cavity also supports mechanical modes that couple back to the polariton fields. Local displacements $x_j(t)$ respond to the polariton densities through radiation$-$pressure$-$like forces enhanced by the excitonic fraction of the polaritons. Their dynamics are governed by
\begin{equation}
\label{eq:xj}
\ddot{x}_j + \Gamma_\mathrm{M} \dot{x}_j + \Omega_\mathrm{M}^2 x_j
= 4 g_j \rho_0 \Omega_\mathrm{M} \mathrm{Re}\left[ \tilde{\Psi}_{j+}\tilde{\Psi}^*_{j-}\right],
\end{equation}
where $\Omega_\mathrm{M}$ and $\Gamma_\mathrm{M}$ denote the mechanical resonance frequency and damping rate. The local mechanical modes primarily modify the dynamics of each individual condensate, affecting the on$-$site spin coupling, $J_j(x_j) = J_{j,0} + \hbar\delta g_0 x_j$. This optomechanical feedback loop strongly influences the oscillation frequencies of the two TCs and reshapes their relative energy landscape.

\begin{figure}[H]
    \centering 
    \includegraphics[trim = 0mm 0mm 0mm 0mm,clip=true, keepaspectratio=true, width=1\columnwidth,angle=0]{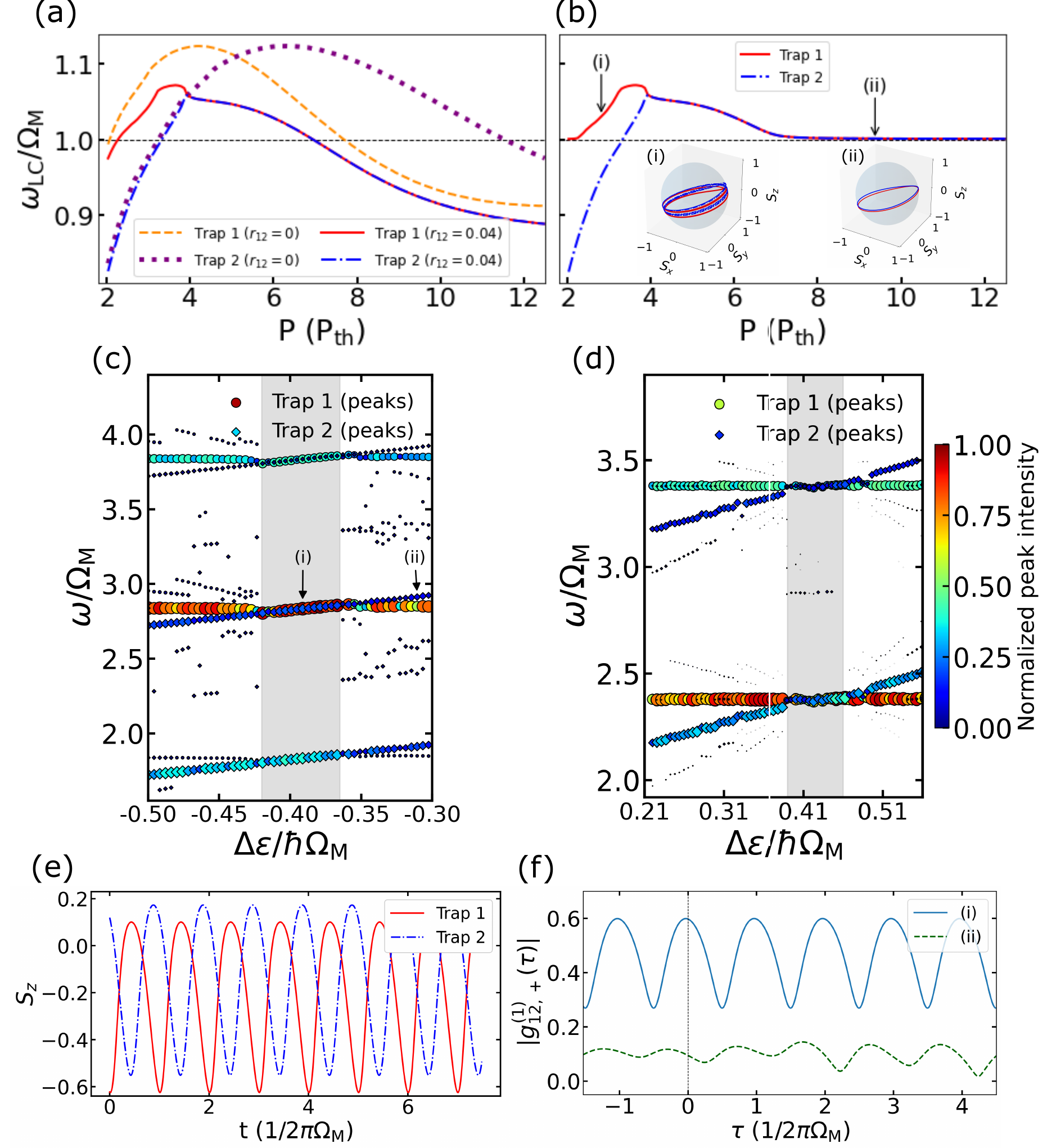}
    \caption{
        \textbf{Coherent and dissipative synchronization of two polariton time crystals.} 
(a,b) Precession frequency of the pseudospin limit cycle in each trap, $\omega_{\mathrm{LC}}$, normalized to the mechanical frequency $\Omega_\mathrm{M}$, as a function of the average pump power $P$, for the pumping conditions and trap parameters given in the main text and Methods. The two curves correspond to the two spatially separated condensates (trap~1 and trap~2). In panel~(a), phonons are absent ($g_1=g_2=0$). In the uncoupled case ($r_{12}=r_{21}=0$), the two traps oscillate at different frequencies, whereas introducing weak crossed-reservoir scattering ($r_{12}=r_{21}=0.02$) causes the two frequencies to progressively converge and merge around $P\!\approx\!4$, signaling dissipative frequency locking and the formation of a synchronized pair of TCs. 
In panel~(b), phonons are allowed to develop in the more strongly pumped trap ($\hbar g_1=3.5\times10^{-5}$). In this case, the oscillation frequencies of both condensates lock to the mechanical frequency, indicating mechanically mediated synchronization. The insets marked (i) and (ii) show the corresponding trajectories of the local pseudospin in each trap on the Poincar\'e sphere, for the parameters indicated by the arrows. Colors and line styles follow the same convention as in the main panel.
(c,d) Frequency spectra obtained from a sweep of the inter-trap detuning $\Delta\varepsilon$, for $\hbar g_1=\hbar g_2=3.5\times10^{-5}$. Panel~(c) corresponds to $P=12.2$, while panel~(d) corresponds to $P=9.4$. Peak positions of the dominant spectral components in each trap are shown, with point size and color encoding the normalized peak intensity. The shaded regions highlight detuning ranges where inter-trap locking occurs. These data demonstrate that synchronization can be induced either by coherent inter-trap coupling ($J_{12}=0.02$, with $r_{12}=r_{21}=0.02$) in panel (c) or by purely dissipative coupling ($J_{12,d}=0.02$) in panel (d).
(e) Time evolution of the $S_{z,j}$ component of the pseudospin ($j=1,2$) for a representative synchronized state, marked as (i) within the shaded region of panel~(c). The red solid line corresponds to trap~1 and the blue dashed line to trap~2, illustrating frequency-locked oscillations of the two local TCs at the phonon frequency $\Omega_{\mathrm{M}}$.
(f) First-order cross-correlation function of the spin-up component between the two traps, $|g^{(1)}_{12,+}(\tau)|$, computed for two representative cases: a synchronized state, marked as (i) in panel~(c), and a non-synchronized state, marked as (ii). The synchronized case exhibits a pronounced oscillatory structure with a maximum close to zero delay, whereas the non-synchronized case shows strongly suppressed correlations. 
}
     \label{Fig4}
\end{figure}

Figure~\ref{Fig4} summarizes the different synchronization mechanisms that emerge from the coupled TC dynamics described by Eqs.~(\ref{eq:psij})--(\ref{eq:xj}). In these simulations, mimicking experimental conditions, the average power $P=\sum_{j,\sigma}P_{j\sigma}/4$ is swept while fixing the relative ratio among the $P_{j\sigma}$, considering a small reservoir polarization of $(P_{j\uparrow}-P_{j\downarrow})/(P_{j\uparrow}+P_{j\downarrow})= 0.2$ and setting trap 1 as the more strongly pumped with $P_{1\sigma}/P_{2\sigma}=1.5$. The remaining intra-trap parameters are provided in Methods. We begin by considering the role of purely dissipative coupling mediated by the exciton reservoirs. Figures~\ref{Fig4}(a,b) show the evolution of the oscillation frequency of the spontaneous limit cycle in each trap as the non-resonant pump power is increased while maintaining fixed relative pump ratios between trap and spin components. The quantity plotted, $\omega_\mathrm{LC}$, corresponds to the precession rate of the condensate pseudospin at each trap, which directly measures the TC frequency arising from the self-sustained nonlinear dynamics of each polariton mode, as established in previous work~\cite{Carraro2024}.

In Fig.~\ref{Fig4}(a), optomechanical interactions are absent ($g_1=g_2=0$). At low pump powers the two traps sustain limit cycles with different intrinsic frequencies, reflecting their distinct local conditions. When weak crossed-reservoir scattering is introduced ($r_{12}=r_{21}=0.02$), the two oscillation frequencies progressively drift toward one another as the pump power, and hence the polariton density, increases. Around $P\!\approx\!4$, the two trajectories merge and the condensates become frequency locked, forming a synchronized pair of TCs. Importantly, in this regime the only interaction between the traps is reservoir mediated; no coherent tunneling or optomechanical coupling is present. The observed locking therefore demonstrates that non-Hermitian coupling alone suffices to synchronize spatially separated polariton TCs.

Figure~\ref{Fig4}(b) illustrates the effect of optomechanical back--action when phonons are allowed to develop in the more strongly pumped trap ($g_1\neq0$, $g_2=0$). In this regime, the oscillation frequencies of both TCs become locked to the mechanical frequency $\Omega_\mathrm{M}$, even though only one trap is directly coupled to the phonon mode. This behavior mirrors the experimentally observed spin--phonon locking shown in Fig.~\ref{Fig1}(c) and highlights the mechanical mode as a common clock for the two spatially separated condensates, imposing a stable and well-defined oscillation frequency. The insets labeled (i) and (ii) show the corresponding trajectories of the local pseudospin in each trap on the Poincar\'e sphere, defined as $S_{j,i}=\tilde{\Psi}_i^{\dagger}\sigma_j\tilde{\Psi}_i$, where $\sigma_j$ are the Pauli matrices ($j=x,y,z$) and $\tilde{\Psi}_i=(\tilde{\Psi}_{i+},\tilde{\Psi}_{i-})^{\mathrm{T}}$. These trajectories reveal the transition from independent oscillations to stable synchronized limit cycles, corresponding to the mechanically locked regime indicated by (ii).

To elucidate the origin of inter-trap synchronization, we next examine the steady-state frequency spectra as a function of the bare inter-trap detuning $\Delta\varepsilon = \hbar ( \omega_2 - \omega_1)$. Figures~\ref{Fig4}(c,d) show the dominant spectral peaks extracted from the pseudospin dynamics, with the color scale and point size encoding the normalized peak intensity. In Fig.~\ref{Fig4}(c), coherent inter-trap coupling ($J_{12}=0.02$) produces a detuning window in which the upper main peak of the TC in the lower-energy trap synchronizes with the lower main peak of the higher-energy trap. This hybrid locking mechanism corresponds directly to the experimental observations reported in Fig.~\ref{Fig2}, where synchronization occurs between distinct energy branches of the two condensates.

Figure~\ref{Fig4}(d) shows the case of purely dissipative inter-trap coupling ($J_{12,d}=0.02$). In this regime, the two main spectral lines of the condensates become fully synchronized over a finite detuning range, reproducing the behavior observed experimentally in Fig.~\ref{Fig3}. Panels~(c) and (d) are designed to mimic the experimental conditions presented in Figs.~\ref{Fig2} and~\ref{Fig3}, respectively. In the former case, panel~(c) demonstrates single-peak synchronization mediated by coherent coupling, as expected for traps whose spatial separation allows for evanescent overlap. Then, panel~(d) shows full synchronization driven by dissipative coupling, consistent with the behavior anticipated for more distant traps where coherent tunneling is suppressed. 

Finally, Figs.~\ref{Fig4}(e,f) provide time-domain and correlation-based signatures of the synchronized regime highlighted in Fig.~\ref{Fig4}(c). Figure~\ref{Fig4}(e) shows the temporal evolution of the circular pseudospin component $S_{z,j}$ in each trap for a representative synchronized state, marked as (i) in panel~(c). The two condensates exhibit frequency-locked oscillations at the mechanical frequency, directly reflecting the emergence of synchronized TC dynamics.

Figure~\ref{Fig4}(f) displays the corresponding first-order cross-correlation function between the two trapped condensates,
\begin{equation}
\left| g^{(1)}_{12,+}(\tau) \right|
=
\left|
\frac{\left\langle \tilde{\Psi}_{1,+}^\ast(t)\,\tilde{\Psi}_{2,+}(t+\tau)\right\rangle_t}
{\sqrt{\left\langle |\tilde{\Psi}_{1,+}(t)|^2\right\rangle_t
\left\langle |\tilde{\Psi}_{2,+}(t)|^2\right\rangle_t}}
\right|,
\end{equation}
where $\tau$ denotes the relative time delay. The correlation function is shown for two representative cases: a synchronized state, labeled (i) in panel~(c), and a non-synchronized state, labeled (ii). In the synchronized regime, $|g^{(1)}_{12,+}(\tau)|$ exhibits a pronounced oscillatory structure with a maximum close to zero delay, signaling strong mutual coherence between the two TCs and phase-locking. This behavior closely matches the experimental cross-correlation measurements reported in Fig.~\ref{Fig2}(c) and provides a clear dynamical signature of TC synchronization. By contrast, the non-synchronized case (ii) shows strongly suppressed correlations. Together, these results confirm that coherent and dissipative interactions, assisted by optomechanical coupling, can robustly synchronize the limit-cycle dynamics of spatially separated polariton condensates.


\section*{Discussion and Outlook}

Coupled spinor exciton--polariton condensates provide a solid-state platform in which CTCs can interact and organize collectively. When individual condensates enter a regime of spontaneous time-translation symmetry breaking, their limit-cycle pseudospin dynamics can synchronize through a combination of coherent, dissipative, and optomechanical coupling mechanisms. The resulting collective states exhibit locked oscillation frequencies and synchronized energies, demonstrating that time-crystalline order can persist and reorganize in genuinely interacting, non-Hermitian quantum systems.

A key finding is that synchronization does not rely exclusively on coherent tunneling. Even in the absence of significant wavefunction overlap, reservoir-mediated dissipative coupling is sufficient to induce energy synchronization and to seed time-crystalline dynamics in a distant condensate. This establishes dissipation as an active resource for stabilizing collective temporal order and provides an experimental realization of TC seeding in a driven-dissipative solid-state system. At shorter separations, coherent Josephson coupling enables additional synchronization channels, selectively hybridizing spectral branches of the two condensates and giving rise to locking scenarios inaccessible in purely conservative settings.

Optomechanical back-action further enriches the dynamics by introducing a long-lived internal clock. Confined hypersound phonons can entrain multiple condensates, enforcing frequency locking of both time crystals even when only one trap couples directly to the mechanical mode. This behavior establishes a direct connection between time-crystalline order and cavity optomechanics, and demonstrates how mechanical degrees of freedom can act as global mediators of temporal coherence.

Beyond frequency locking, coupled polariton TCs self-organize into dynamical ferro- and antiferromagnetic configurations, defined by the relative orientation of their rotating pseudospins. Unlike previously studied static spin-ordered phases, these states correspond to collective limit-cycle attractors, where ordering occurs in time rather than in stationary observables. The ability to reversibly switch between these configurations using optical power underscores the high tunability of the platform.

These results extend the notion of time-translation symmetry breaking from isolated oscillators to interacting networks of non-Hermitian quantum condensates, linking CTCs to broader concepts of synchronization and collective dynamics. Scaling to larger trap arrays, engineering programmable dissipative couplings, and exploiting hybrid phononic--temporal architectures offer natural routes to explore complex spatiotemporal phases and collective time-crystalline matter beyond equilibrium paradigms. The proposed system could be used as on chip ultra-precise self-synchronizing optical clocks operating above 100 GHz.

\section*{Methods}
\label{sc:Meth}
\subsection*{Microcavity sample and polariton traps}

The experiments were performed on a patterned (Al,Ga)As planar microcavity operating in the strong light--matter coupling regime. The sample structure and fabrication procedure are identical to those described in the Supplementary Material of Ref.~\cite{Carraro2024}, and are summarized here for completeness.

The microcavity consists of two distributed Bragg reflectors (DBRs) enclosing a $3/2\lambda$-thick AlAs spacer embedding four 15-nm-thick GaAs quantum wells (QWs). The QWs are positioned slightly displaced from the antinodes of the cavity electric field so as to maximize optomechanical coupling while preserving strong exciton--photon coupling~\cite{Zambon2022,Sesin2023}. The lower (upper) DBR comprises 33 (25) $\lambda/4$ periods of Al$_{0.15}$Ga$_{0.85}$As/Al$_{0.9}$Ga$_{0.1}$As layers.

Lateral confinement of polaritons was realized using an etch-and-overgrowth technique applied to the cavity spacer between molecular beam epitaxy growth steps. Shallow mesas of a few nanometers in height were defined by photolithography and wet chemical etching, inducing a local blueshift of the cavity photon mode in the etched regions while leaving the quantum wells unaffected. Subsequent overgrowth restores a planar surface while preserving the lateral potential landscape.

In addition to optical confinement, the cavity is engineered to act as an efficient optomechanical resonator. As previously shown, a so-called ``magic coincidence'' exists for this material system and cavity design, resulting in simultaneous confinement of near-infrared photons and hypersound phonons~\cite{Fainstein2013}. The confined mechanical modes correspond to longitudinal breathing vibrations polarized along the growth direction, with discrete frequencies
\[
\nu^{(n)} = (2n+1)\times 20~\mathrm{GHz},
\]
where $n$ is an integer. The spacer thickness and QW positions are chosen such that the exciton-mediated deformation potential interaction is optimized: the QWs are displaced from the antinodes of the cavity electric field $E(z)$, which coincide with nodes of the strain field $s(z)$, towards positions where the product $s(z)\lvert E(z)\rvert^{2}$ is maximized. This design leads to concurrent confinement of phonons and photons within the cavity and maximizes their mutual coupling. Hypersound reflection measurements yield a linewidth of the confined phonon modes of $\Gamma \sim 3$~MHz at 10~K, corresponding to phonon lifetimes several orders of magnitude longer than the coherence time of the polariton condensates.

Three different trap configurations were investigated. The first configuration consists of two square traps with lateral size $4\times4~\mu\mathrm{m}^2$ separated by a $1~\mu\mathrm{m}$-wide barrier (Fig.~1). The second configuration consists of two square traps with lateral size $2\times2~\mu\mathrm{m}^2$ separated by a $2~\mu\mathrm{m}$-wide barrier (Fig.~2). The third configuration consists of two $2\times2~\mu\mathrm{m}^2$ traps separated by a $2~\mu\mathrm{m}$ barrier, together with an additional identical trap located $13~\mu\mathrm{m}$ away from the center of the right trap (Fig.~3).

All measurements were performed at a negative photon--exciton energy detuning in the range $-10$ to $-7$~meV in the non-etched regions, corresponding to polaritons with a predominantly photonic character.

\subsection*{Optical excitation and photoluminescence measurements}

Experiments were carried out in a closed-cycle helium cryostat at a temperature of $\sim 5$~K. Polaritons were excited using a continuous-wave single-mode Ti:sapphire laser operating at $760$~nm ($1.631$~eV), well above the exciton resonance, ensuring non-resonant excitation via a hot excitonic reservoir.

The excitation beam was focused onto the sample at normal incidence using a microscope objective (typically $\times20$, NA $=0.3$), resulting in a Gaussian spot with a diameter of approximately $3~\mu$m. For measurements on coupled traps, the excitation spot was typically positioned closer to one of the two traps in order to introduce a controlled asymmetry in the reservoir feeding. The excitation power was varied to access the condensation regime and the time-dependent dynamical states reported in this work.

The polarization state of the excitation beam was controlled using a quarter-wave plate placed before the cryostat entrance window. By rotating this wave plate, the ellipticity of the pump polarization was continuously tuned, allowing to have control over the spin polarization (magnetization) of the excitonic reservoir generated by the non-resonant excitation.

Photoluminescence (PL) from the traps was collected along the sample normal and analyzed using a micro-photoluminescence setup. The real-space image of the sample was projected onto the entrance slit of the spectrometer, providing spatial resolution along one dimension and enabling unambiguous identification of the emission originating from each individual trap. Standard spectral measurements were performed using a triple-additive spectrometer providing a spectral resolution of approximately $0.01$~meV ($\sim 2.4$~GHz) and recorded with a liquid-nitrogen-cooled CCD camera. This resolution was sufficient to resolve narrow emission lines and modulation-induced spectral features associated with the pseudospin precession of the polariton condensates. A notch filter centered at the excitation wavelength was used to suppress scattered pump light.

Polarization-resolved PL detection was implemented using a sequence of polarization optics placed before the entrance of the spectrometer. First, a quarter-wave plate converted the circularly polarized components of the polariton emission into orthogonal linear polarizations (diagonal and anti-diagonal for right- and left-circular polarization, respectively). A half-wave plate placed after the quarter-wave plate was then used to rotate the linear polarization basis. Finally, the emission passed through a vertically oriented linear polarizer. By setting the half-wave plate angle to $+22.5^\circ$ or $-22.5^\circ$, the detection path selectively transmitted either the right- or left-circularly polarized component of the polariton emission.

\subsection*{First-order coherence measurements}

Temporal coherence and dynamical properties were investigated by measuring the first-order correlation function $g^{(1)}(r,\tau)$ using a modified Michelson interferometer. The collected photoluminescence (PL) was split by a non-polarizing beam splitter and sent into the two interferometer arms: one of fixed length and the other incorporating a motorized retro-reflector acting as a variable delay line. The two beams were recombined at a small non-zero relative angle and imaged onto a CCD camera, producing spatial interference fringes. The modulus of the first-order coherence function,
\begin{equation}
g^{(1)}(r,\tau)=
\frac{\langle \mathbf{E}^\ast(r,t)\,\cdot\mathbf{E}(r,t+\tau)\rangle}
{\sqrt{\langle |\mathbf{E}(r,t)|^2\rangle \langle |\mathbf{E}(r,t+\tau)|^2\rangle}},
\end{equation}
where $\mathbf{E}(r,t)$ is the complex electric field and the brackets denote temporal averaging, is directly encoded in the fringe visibility. The complex coherence field was extracted from the interferograms using the Takeda method, in which the spatial interferogram is Fourier-transformed, one of the first-order diffraction sidebands is isolated by spectral filtering, and an inverse Fourier transform yields the complex interference term~\cite{Takeda2005}. From this procedure, the spatially resolved coherence amplitude $|g^{(1)}(r,\tau)|$ was obtained and subsequently integrated over the spatial coordinate $r$ to yield the delay-dependent coherence function $|g^{(1)}(\tau)|$.

For autocorrelation measurements, the emission originating from a given trap was interfered with itself, and the delay-dependent fringe visibility was extracted as a function of the optical path difference between the two arms.

For cross-correlation measurements between different traps, the real-space image of one trap was optically superimposed onto that of another trap at the entrance of the interferometer while preserving independent emission paths for the two condensates. By scanning the delay line, the mutual first-order coherence between spatially separated polariton condensates was directly measured. This configuration allows access to both the relative phase and the temporal dynamic between coupled and distant polariton condensates.

Further details of the interferometric setup and the data analysis procedures are provided in the Supplementary Information of Ref.~\cite{Carraro2024}.

\subsection*{Numerical simulation details} 

We solve Eqs.~\eqref{eq:psij} and~\eqref{eq:xj} taking identical intra-trap parameters for both traps and using the mechanical frequency $\Omega_\mathrm{M}=1$ as the energy scale. In dimensionless units, the non-linear interaction strengths are set to $U_0 = 0.1$ and $U_0^\mathrm{R} = 1.7$, while the coherent intra-trap spin-mixing term is $J_{j,0}=0.165$. The polariton and reservoir decay rates are $\gamma=0.26$ and $\gamma_\mathrm{R}=0.23$, respectively, and the characteristic density scale is $\rho_0=2500$. 

When phonon--polariton interactions are included, the cavity phonons in both traps are characterized by a damping rate $\Gamma_\mathrm{M} = 2.5\times10^{-4}$, corresponding to a mechanical quality factor $Q_\mathrm{M}=4000$, and an optomechanical coupling strength $\hbar g_j = 3.5\times10^{-5}$.

\subsection*{Data availability}

The data that support the findings of this study are included within the main text and Methods and are available from the corresponding author upon reasonable request.


\begin{acknowledgments}
We acknowledge partial financial support from the ANPCyT-FONCyT (Argentina) under grants PICT-2018-03255, PICT 2018-1509, PICT 2019-0371, PICT 2020-3285, and SECTyP UNCuyo 06/C053-T1.
ASK and PVS acknowledge the funding from German DFG (grant 359162958).  
\end{acknowledgments}


\begin{thebibliography}{40}


\bibitem{Wilczek2012} F. Wilczek, Quantum time crystals, Phys. Rev. Lett. \textbf{109}, 160401 (2012).

\bibitem{Bruno2013} P. Bruno, Impossibility of spontaneously rotating time crystals: a no-go theorem, Phys. Rev. Lett. \textbf{111}, 070402 (2013).

\bibitem{Sacha2018} K. Sacha, and J. Zakrzewski, Time crystals: a review, Rep. Prog. Phys. \textbf{81}, 016401 (2018).


\bibitem{Kessler2021} H. Kessler, P. Kongkhambut, C. Georges, L. Mathey, J. G. Cosme , and A. Hemmerich,  Observation of a Dissipative Time Crystal, Phys. Rev. Lett. \textbf{127}, 043602 (2021).


\bibitem{Autti2018} S. Autti, V. B. Eltsov, and G. E. Volovik, Observation of a Time Quasicrystal and Its Transition to a Superfluid Time Crystal, Phys. Rev. Lett. \textbf{120}, 215301 (2018).



\bibitem{Gong2018} Z. Gong, R. Hamazaki, and M. Ueda, Discrete Time-Crystalline Order in Cavity and Circuit QED Systems, Phys. Rev. Lett. \textbf{120}, 040404 (2018).

\bibitem{Tucker2018} K. Tucker, B. Zhu, R. J. Lewis-Swan, J. Marino, F. Jimenez, J. G. Restrepo, and A.M. Rey, Shattered time: Can a dissipative time crystal survive many-body correlations? New J. Phys. \textbf{20}, 123003 (2018).

\bibitem{Zhu2019} B. Zhu, J. Marino, N. Y. Yao, M. D. Lukin, and E. A. Demler, Dicke time crystals in driven-dissipative quantum many-body systems, New J. Phys. 21, 073028 (2019).

\bibitem{Buca2019} B. Buca, J. Tindall, and D. Jaksch, Non-stationary coherent quantum many-body dynamics through dissipation, 
Nature Communications \textbf{10}, 1730 (2019).

\bibitem{Lazarides2020} A. Lazarides, S. Roy, F. Piazza, and R. Moessner, R. Time crystallinity in dissipative Floquet systems, Phys. Rev. Res. \textbf{2}, 022002 (2020).

\bibitem{Else2020} D. V. Else, C. Monroe, C. Nayak, and N. Y. Yao, Discrete time crystals, Annu. Rev. Condens. Matter Phys. \textbf{11}, 467 (2020).



\bibitem{Zhang2017} J. Zhang, P. W. Hess, A. Kyprianidis, P. Becker, A. Lee, J. Smith, G. Pagano, I.-D. Potirniche, A. C. Potter, A. Vishwanath, N. Y. Yao, and C. Monroe, Observation of a discrete time crystal, Nature \textbf{543}, 217 (2017).


\bibitem{Taheri2022} H. Taheri, A. B. Matsko, L. Maleki, and Krzysztof Sacha, All-optical dissipative discrete time crystals, Nat. Comm. \textbf{13}, 848 (2022).


\bibitem{Frey2022} P. Frey, and S. Rachel, Realization of a discrete time crystal on 57 qubits of a quantum computer, Sci. Adv. \textbf{8}, eabm7652 (2022). 



\bibitem{Kessler2019} H. Kessler, J. G. Cosme, M. Hemmerling, L. Mathey, and A. Hemmerich, Emergent limit cycles and time crystal dynamics in an atom-cavity system, Phys. Rev. A \textbf{99}, 053605 (2019). 

\bibitem{Kongkhambut2022} P. Kongkhambut, J. Skulte, L. Mathey, J. G. Cosme, A. Hemmerich, and H. Kessler, Observation of a continuous time crystal, Science \textbf{377}, 670 (2022).


\bibitem{Greilich2024} A. Greilich, N. E. Kopteva, A. N. Kamenskii, P. S. Sokolov, V. L. Korenev, and M. Bayer, Robust continuous time crystal in an electron-nuclear spin system, Nature Physics \textbf{20}, 631 (2024).

\bibitem{Carraro2024} I. Carraro-Haddad, D. L. Chafatinos , A. S. Kuznetsov, I. A. Papuccio-Fern\'andez , A. A. Reynoso , A. Bruchhausen, K. Biermann, P. V. Santos, G. Usaj, and A. Fainstein, Solid-state continuous time crystal in a polariton condensate with a built-in mechanical clock, Science \textbf{384}, 995 (2024).



\bibitem{Iemini2018} F. Iemini, A. Russomanno, J. Keeling, M. Schir\'o, M. Dalmonte, and R. Fazio, Boundary Time Crystals, Phys. Rev. Lett. \textbf{121}, 035301 (2018).

\bibitem{Hajdusek2022} M. Hajdu\^sek, P. Solanki, R. Fazio, and S. Vinjanampathy, 
Seeding Crystallization in Time, Phys. Rev. Lett. \textbf{128}, 080603 (2022).

\bibitem{Schiro2016} M. Schir\'o, C. Joshi, M. Bordyuh, R. Fazio,J. Keeling, and H. E. T\"ureci, 
Exotic Attractors of the Nonequilibrium Rabi-Hubbard Model, Phys. Rev. Lett. \textbf{116}, 143603 (2016).

\bibitem{Autti2021} S. Autti, P. J. Heikkinen, J. T. M\"akinen, G. E. Volovik, V. V. Zavjalov, and V. B. Eltsov, AC Josephson effect between two superfluid time crystals, Nature Mat.\textbf{ 20}, 171 (2021).






\bibitem{Hartmann2006} M. J. Hartmann, F. G. Brandao, and M. B. Plenio, Strongly interacting polaritons in coupled arrays of cavities, Nature Physics \textbf{2}, 849 (2006).

\bibitem{Amo2009} A. Amo, D. B. D. Sanvitto, F. Laussy, E. del Valle, M. Martin, A. Lema\^{\i}tre, J. Bloch, D. Krizhanovskii, M. Skolnick, C. Tejedor, and L. Vi\~na, Collective Fluid Dynamics of a Polariton Condensate in a Semiconductor Microcavity, Nature (London) \textbf{457}, 291 (2009).

\bibitem{CarusottoRMP2013} I. Carusotto, and C. Ciuti, Quantum fluids of light, Reviews of Modern Physics \textbf{85}, 299 (2013). 

\bibitem{Rozas2014} G. Rozas, A. E. Bruchhausen, A. Fainstein, B. Jusserand, and A. Lema\^itre, 
Polariton path to fully resonant dispersive coupling in optomechanical resonators, 
Phys. Rev. B \textbf{90}, 201302(R) (2014).


\bibitem{Kasprzak2006} J. Kasprzak, M. Richard, S. Kundermann, A. Baas, P. Jeambrun, J. M. J. Keeling, F. M. Marchetti, M. H. Szyma\'nska, R. Andr\'e, J. L. Staehli, V. Savona, P. B. Littlewood, B. Deveaud, and L. S. Dang, Bose-Einstein Condensation of Exciton Polaritons, Nature (London) \textbf{443}, 409 (2006).

\bibitem{Balili2007} R. Balili, V. Hartwell, D. Snoke , L. Pfeiffer, and K. West, Bose-Einstein Condensation of Microcavity Polaritons in a Trap, Science \textbf{316}, 1007 (2007).

%
%
\bibitem{Schneider2016} C. Schneider, K. Winkler, M. D. Fraser, M. Kamp, Y. Yamamoto, E. A. Ostrovskaya, and S. H\"ofling, Exciton-polariton trapping and potential landscape engineering, Rep. Prog. Phys. \textbf{80}, 016503 (2016).
%
%
\bibitem{Bajoni2008} D. Bajoni, P. Senellart, E. Wertz, I. Sagnes, A. Miard, A. Lema\^itre, and J. Bloch,  Polariton Laser Using Single Micropillar  GaAs-GaAlAs, Semiconductor CavitiesPhys. Rev. Lett. \textbf{100}, 047401 (2008).

\bibitem{Galbiati2012}  M. Galbiati, L. Ferrier, D. D. Solnyshkov, D. Tanese, E. Wertz, A. Amo, M. Abbarchi, P. Senellart, I. Sagnes, A. Lema\^itre, E. Galopin, G. Malpuech, and J. Bloch, Polariton Condensation in Photonic Molecules, Phys. Rev. Lett. \textbf{108}, 126403 (2012).
%
%
\bibitem{Askitopoulos2013} A. Askitopoulos, H. Ohadi, A. V. Kavokin, Z. Hatzopoulos, P. G. Savvidis, and P. G. Lagoudakis,  Polariton condensation in an optically induced two-dimensional potential, Phys. Rev. B \textbf{88}, 041308(R) (2013).

\bibitem{Cristofolini2013} P. Cristofolini, A. Dreismann, G. Christmann, G. Franchetti, N. G. Berloff, P. Tsotsis, Z. Hatzopoulos, P. G. Savvidis, and J. J. Baumberg,  Optical Superfluid Phase Transitions and Trapping of Polariton Condensates, Phys. Rev. Lett. \textbf{110}, 186403 (2013).

\bibitem{Alyatkin2021} S. Alyatkin, H. Sigurdsson, A. Askitopoulos, J. D. T\"opfer, and P. G. Lagoudakis,  Quantum fluids of light in all-optical scatterer lattices, Nature Communications  \textbf{12}, 5571 (2021).
%
%
\bibitem{Winkler2015}  K. Winkler, J. Fischer, A. Schade, M. Amthor, Robert Dall, Jonas Gessler, M. Emmerling, E. A. Ostrovskaya, M. Kamp, C. Schneider, and S. H\"ofling,  A polariton condensate in a photonic crystal potential landscape, New J. Phys. \textbf{17}, 023001 (2015).

\bibitem{Kuznetsov2018} A. S. Kuznetsov, P. L. J. Helgers, K. Biermann, and P. V. Santos, Quantum Confinement of Exciton-Polaritons in Structured (Al,Ga)As Microcavity, Phys. Rev. B \textbf{97}, 195309 (2018).



\bibitem{Fainstein2013} A. Fainstein, N. D. Lanzillotti-Kimura, B. Jusserand, B. Perrin, Strong optical-mechanical coupling in a vertical GaAs/AlAs microcavity for subterahertz phonons and near-infrared light, Physical Review Letters \textbf{110}, 037403 (2013).

\bibitem{Santos2023} P. V. Santos and A. Fainstein, Polaromechanics: polaritonics meets optomechanics, Optical Materials Express \textbf{13}, 1974 (2023).


\bibitem{RMP} M. Aspelmeyer, T. J. Kippenberg, and F. Marquardt, Cavity Optomechanics, Rev. Mod. Phys. \textbf{86}, 1391 (2014).


\bibitem{Chafatinos2020} D. L. Chafatinos, A. S. Kuznetsov, S. Anguiano, A. E. Bruchhausen, A. A. Reynoso, K. Biermann, P. V. Santos, and A. Fainstein, Polariton-driven phonon laser,  Nature Communications \textbf{11}, 4552 (2020).

\bibitem{Reynoso2022} A. A. Reynoso, G. Usaj, D. L. Chafatinos, F. Mangussi, A. E. Bruchhausen, A. S. Kuznetsov, K. Biermann, P. V. Santos, and A. Fainstein, Optomechanical parametric oscillation of a quantum light-fluid in a lattice, Phys. Rev. B \textbf{105}, 195310 (2022).

\bibitem{Chafatinos2023}D. L. Chafatinos, A. S. Kuznetsov, P. Sesin, I. Papuccio, A. A. Reynoso, A. E. Bruchhausen, G. Usaj, K. Biermann, P V. Santos, A. Fainstein, Asynchronous Locking in Metamaterials of Fluids of Light and Sound, Nature Communications \textbf{14}, 3485 (2023).

\bibitem{Ramos2024} I. A. Ramos-P\'erez, I. Carraro-Haddad, F. Fainstein, D. L. Chafatinos, G. Usaj, G. B. Mindlin, A. Fainstein, and A. A. Reynoso, Theory of optomechanical locking in driven-dissipative coupled polariton condensates, Phys. Rev. B \textbf{109}, 165305 (2024).







\bibitem{Nalitov2019} A. V. Nalitov, H. Sigurdsson, S. Morina, Y. S. Krivosenko, I. V. Iorsh, Y. G. Rubo, A. V. Kavokin, and I. A. Shelykh, Optically trapped polariton condensates as semiclassical time crystals, Phys. Rev. B \textbf{99}, 033830 (2019).




\bibitem{Wouters2007} M. Wouters and I. Carusotto, Excitations in a Nonequilibrium Bose-Einstein Condensate of Exciton Polaritons, Phys. Rev. Lett. \textbf{99}, 140402 (2007).


\bibitem{Shelykh2005} I. A. Shelykh, A. V. Kavokin, and G. Malpuech, Spin dynamics of exciton polaritons in microcavities, phys. stat. sol. (b) 242, 2271 (2005).


\bibitem{Ohadi2015} H. Ohadi, A. Dreismann, Y. G. Rubo, F. Pinsker, Y. del Valle-Inclan Redondo, S. I. Tsintzos, Z. Hatzopoulos, P. G. Savvidis, and J. J. Baumberg, Spontaneous Spin Bifurcations and Ferromagnetic Phase Transitions in a Spinor Exciton-Polariton Condensate, Phys. Rev. X \textbf{5}, 031002 (2015).

\bibitem{Gnusov2020} I. Gnusov, H. Sigurdsson , S. Baryshev, T. Ermatov, A. Askitopoulos, and P. G. Lagoudakis, Optical orientation, polarization pinning, and depolarization dynamics in optically confined polariton condensates, Phys. Rev. B \textbf{102}, 125419 (2020).

\bibitem{Siggurdsson2020} H. Sigurdsson, Hysteresis in linearly polarized nonresonantly driven exciton-polariton condensates, Phys. Rev. Res. \textbf{2}, 023323 (2020). 


\bibitem{Ohadi2017} H. Ohadi, A. J. Ramsay, H. Sigurdsson, Y. del Valle-Inclan Redondo, S. I. Tsintzos, Z. Hatzopoulos, T. C. H. Liew, I. A. Shelykh, Y. G. Rubo, P.G. Savvidis, and J. J. Baumberg, Spin Order and Phase Transitions in Chains of Polariton Condensates, Phys. Rev. Lett.textbf{119}, 067401 (2017).

\bibitem{Sigurdsson2017} H. Sigurdsson, A. J. Ramsay, H. Ohadi, Y. G. Rubo, T. C. H. Liew, J. J. Baumberg, and I. A. Shelykh, Driven-dissipative spin chain model based on exciton-polariton condensates, Phys. Rev. B \textbf{96}, 155403 (2017).


\bibitem{Rayanov2015} K. Rayanov, B. L. Altshuler, Y. G. Rubo, and S. Flach, Frequency Combs with Weakly Lasing Exciton-Polariton Condensates, Phys. Rev. Lett. \textbf{114}, 193901 (2015).



\bibitem{Wouters2008} M. Wouters, Synchronized and desynchronized phases of coupled nonequilibrium exciton-polariton condensates, Physical Review B \textbf{77}, 121302 (2008).



\bibitem{Aleiner2012} I. L. Aleiner, B. L. Altshuler, and Y. G. Rubo, Radiative coupling and weak lasing of exciton-polariton condensates,  Phys. Rev. B \textbf{85}, 121301(R) (2012).


\bibitem{Zambon2022} N. Carlon Zambon, Z. Denis, R. De Oliveira, S. Ravets, C. Ciuti, I. Favero, and J. Bloch, Enhanced Cavity Optomechanics with Quantum-Well Exciton Polaritons, Phys. Rev. Lett. \textbf{129}, 093603 (2022).

\bibitem{Sesin2023}  P. Sesin, A. S. Kuznetsov, G. Rozas, S. Anguiano, A. E. Bruchhausen, A. Lema\^itre, K. Biermann, P. V. Santos, and A. Fainstein, Giant optomechanical coupling and dephasing protection with cavity exciton-polaritons, Phys. Rev. Research \textbf{5}, L042035 (2023).


\bibitem{Takeda2005}
M. Takeda, W. Wang, Z. Duan, and Y. Miyamoto,
Coherence holography,
Optics Express \textbf{13}, 9629--9635 (2005).











%
%
%
%

%
%



%





\end{thebibliography}
\end{document}